\documentclass[%
 reprint,
 amsmath,amssymb,
 aps,
 pra,
]{revtex4-2}
\usepackage{amsthm}
\usepackage{graphicx}
\usepackage{dcolumn}
\usepackage{bm,color}

\definecolor{e-mail}{rgb}{0,.40,.80}
\definecolor{reference}{rgb}{.20,.60,.22}
\definecolor{citation}{rgb}{0,.40,.80}

\theoremstyle{theorem}
\newtheorem{lemma}{Lemma}[section]
\newtheorem{example}{Example}[section]
\newtheorem{corollary}{Corollary}[section]

\theoremstyle{definition}
\newtheorem{remark}{Remark}[section]

\usepackage[colorlinks, linkcolor=reference,
            citecolor=citation,
          urlcolor=e-mail]{hyperref}

\DeclareMathOperator{\diag}{diag}
\DeclareMathOperator{\GL}{GL}

\begin{document}

\preprint{APS/123-QED}

\title{Universality of Photonic Interlacing Architectures for Learning Discrete Linear Unitaries}

\author{M. Markowitz}
\author{M.-A. Miri}
\email{mmiri@qc.cuny.edu}
\author{A. Ovchinnikov}
\author{K. Zelaya}
\thanks{The authors are listed in alphabetic order of their last names and share the same credit.}
\affiliation{%
Department of Physics, Queens College of the City University of New York, Queens, New York 11367, USA\\
Department of Mathematics, Queens College of the City University of New York, Queens, New York 11367, USA\\
Mathematics and Computer Science Programs, The Graduate Center, City University of New York, New York, NY 10016, USA\\
Physics Program, The Graduate Center, City University of New York, New York, New York 10016, USA
}%

\date{\today}

\begin{abstract}
Recent investigations suggest that the discrete linear unitary group $U(N)$ can be represented by interlacing a finite sequence of diagonal phase operations with an intervening unitary operator. However, despite rigorous numerical justifications, no formal proof has been provided. Here, we show that elements of $U(N)$ can be decomposed into a sequence of $N$-parameter phases alternating with $1$-parameter propagators of a lattice Hamiltonian. The proof is based on building a Lie group by alternating these two operators and showing its completeness to represent $U(N)$ for a finite number of layers, which is numerically found to be exactly $N$. This architecture can be implemented using elementary optical components and can successfully reconstruct arbitrary unitary matrices. We propose example devices such as optical logic gates, which perform logic gate operations using a single-layer lossless and passive optical circuit design.
\end{abstract}

\maketitle


\section{Introduction}

Since the pioneering work of Reck \textit{et al.}~\cite{reck1994experimental}, there has been an ever-increasing interest in realizing optical structures that perform arbitrary discrete linear unitary operations. This interest is driven by the exciting proprieties of optical signal processing, such as processing capabilities, lower energy consumption, and faster processing speeds than their electronic counterparts. The latter is central in classical and quantum optical information processing~\cite{harris2018linear, Bogaerts20a, Bogaerts20b,harris2017quantum,pelucchi2022potential,luo2023recent}. The rapid developments of photonic integrated circuit technologies have positioned them as a natural platform for large-scale implementation of photonic unitary circuits through multipath interference networks and reconfigurable phase shifters~\cite{clements2016optimal,pastor2021arbitrary,Tang2021, taballione2021,Markowitz23,Markowitz2023auto,zelaya2024goldilocks}. In the context of classical photonics, the implementation of linear operations have gone beyond the realm of unitaries to cover general matrix operations~\cite{miller2013self,markowitz2024learning}, which find applications in optical convolutional units~\cite{meng2023compact,zelaya2024photonic} and orthogonal communication channels~\cite{miller2000communicating,miller2019waves}.

Recent studies have shown promising results in this field, and further research is underway to advance these technologies. Research has been focused on developing devices that can perform all-optical unitary operations packaged in on-chip form-factor architectures~\cite{Tanomura22a,Bogaerts20a,Markowitz23,Markowitz2023auto,zelaya2024goldilocks,Taguchi23,harris2018linear,miller2013self}. Alternative architectures based on cascading of a fixed intervening operator with diagonal phase shift layers, capable of representing universal unitary matrices, have been reported in the literature \cite{pastor2021arbitrary, Tanomura20, Tanomura22a, Tanomura23, saygin_robust_2020, Skryabin2021, Markowitz23}. Such an architecture can be realized on-chip with multimode interference couplers \cite{pastor2021arbitrary}, or multicore waveguide couplers \cite{Tanomura22a, Markowitz23}, interlaced with programmable phase shifters. In particular, recently, we showed that nonuniform photonic lattices of particularly designed coupling coefficients and length to implement a Discrete Fractional Fourier Transform (DFrFT) operation can be utilized as the intervening operation for realizing programmable unitaries through such an interlacing architecture \cite{Markowitz23,Markowitz2023auto}. While a formal proof of the universality of this construction is not currently available, strong numerical evidence, i.e., a phase transition in the norm of representation error, suggests that arbitrary unitary matrices can be realized with remarkable precision, even within the numerical noise error.
 
This work focuses on the factorization of $N \times N$ unitary matrices in terms of diagonal unitaries and families of one-parameter unitaries established through the propagator of lattice Hamiltonians. This approach has two fundamental benefits. Firstly, the one-parameter unitaries enable a group structure that imposes a finite upper limit on the number of matrices necessary for the factorization. Secondly, the proposed factorization can be implemented using elementary optical components, such as phase shifters and waveguide arrays, making it ideal for photonic applications. The so-constructed architecture is numerically tested and shown to reconstruct arbitrary unitary matrices successfully with good accuracy. This is achieved by optimizing the space of parameters defining the architecture, composed of $N^2$ phase shifters and $N$ waveguide lengths, which over-parameterize the optimization problem in question ($N^2$ parameters required for $N\times N$ unitaries). The fact that $N-1$ waveguide lengths are optimized implies that the architecture is not programmable, for it cannot be changed once manufactured. Still, there are instances in which a passive unitary device is required for a multipurpose application, such as beam-steering and direction-finding tasks. In such cases, the current architecture provides a non-tunable solution that requires fewer layers than other approaches~\cite{zelaya2024goldilocks,Markowitz2023auto}. An all-logical logic gate device is proposed to illustrate the applicability of a passive photonic unit. Indeed, such a task has been reported on combined gain and loss in an optical fiber~\cite{williams2013all} and stub optical lattices showcasing a flat band in the dispersion relations~\cite{real2017flat}. However, a non-unitary process is required to account for a lossy system in the former case, whereas the second approach involves a multi-stacked waveguide design. Here, logic gate operations are achieved using a lossless single-stack optical circuit design. This renders an energy-efficient architecture that utilizes four active phase shifters to generate the optical signals in the input ports and simultaneously switches the device logic operation. Such optical signals are then passively processed by the proposed architecture. This provides an architecture design based on silicon-on-silica platforms that can be fabricated in open-access foundries without requiring any specialized fabrication process.
 
The paper is organized as follows. Section~\ref{sec:math} proposes a representation for arbitrary $N\times N$ unitary matrices based on a factorization in terms of other particular unitary matrices. This is essential as it allows for proof of the finiteness of the architecture factorization, yielding a termination criterion for iterative algorithms that search for the optimal factorization length. The latter is corroborated by numerical tests, where a significant drop in the reconstruction error is found at exactly $N$ layers. In Section~\ref{sec:phot}, an all-optical implementation of the proposed factorization is discussed. This allows for a compact on-chip device based merely on coupled waveguides of different lengths and layers of passive phase elements. Particularly, a three-port logic gate photonic circuit is introduced to illustrate the potential applications of the device, which is further reinforced by comparing the ideal model with full-wave simulations. Such a circuit operates by purely passive means, rendering an energy-efficient solution that performs the \texttt{AND}, \texttt{OR}, \texttt{NAND}, and \texttt{XOR} operations.

\section{Proof of Uiversality}
\label{sec:math}
\subsection{Notation}
Let us consider $\mathcal{U}$ as an arbitrary element of the group $U(N)$ of unitary $N\times N$ matrices, $\mathcal{U}\in U(N)$. It is desirable to find a factorization of unitary matrices in terms of a set of unitary matrices that serve as building blocks to reconstruct any $\mathcal{U}\in U(N)$. Indeed, several such factorizations have been introduced and discussed in the literature of lossless photonic architectures~\cite{Markowitz23,Markowitz2023auto,tanomura_robust_2020,saygin_robust_2020}. Throughout this manuscript, we consider the particular factorization,
\begin{equation}
\label{eq_U}
    \mathcal{U} (\boldsymbol{x}) = e^{i Q^{(M)}} e^{i \ell_{M-1} H} e^{i Q^{(M-1)}} \cdots e^{i \ell_{1} H} e^{i Q^{(1)}} ,
\end{equation}
where $M\in\mathbb{Z}^{+}$ is the total number of layers required in the factorization, and $Q^{(n)}$ are diagonal matrices containing the phase components $Q_{j,k}^{(n)}=\phi_{j}^{(n)}\delta_{j,k}$, with $\phi_{j}^{(n)}\in(0,2\pi]$ for all $j,k\in\{1,\ldots,N\}$ and $n\in\{1,\ldots,M\}$. Furthermore, $\boldsymbol{x}$ stands for the tuple composed of the previous parameter $\phi$'s and $\ell$. 

Here, $H=H^{\dagger}$ is a Hermitian matrix in $\mathbb{C}^{N}$ characterizing the coupled waveguide array, whereas $\ell_{n}$ stands for the waveguide coupling length. Since $H$ contains the information about the mode coupling between nearest waveguides, we focus on operators whose matrix representation renders a thee-diagonal matrix; i.e., 
\begin{equation}\label{eq:Hmatrix}
H_{p,q}=\kappa_{p-1}\delta_{p,q-1}+\kappa_{p}\delta_{p,q+1}+\nu_{p}\delta_{p,q},
\end{equation}
for $p,q,\in\{1,\ldots,N-1\}$. This structure follows from the coupled-mode theory~\cite{Huang94} and the evanescent wave coupling between nearest neighbors and onsite coupling due to phase mismatch. 
From~\eqref{eq_U}, it is clear that the architecture contains $NM$ programmable phase shifter parameters and $N-1$ coupling lengths to be optimized based on the given target matrix $\mathcal{U}$. Remark that the length of the waveguide arrays is fixed and cannot be adjusted within this particular architecture, which can be optimized beforehand in order to construct the unitary optical unit of interest. The mathematical proof below ensures that any unitary operator can be constructed this way. 
\subsection{Existence of layer number bound}\label{sec:exbound}
In this section, we prove that, for every positive $N$, any $\varepsilon >0$, and any unitary $N\times N$ matrix, there exists $M$ and a factorization~\eqref{eq_U} of length $M$ for $\mathcal{U}$ up to $\varepsilon$-error ($\varepsilon = 0$ for a subclass of cases, explained below). This result serves as a termination criterion for a numerical (e.g., optimization-based) algorithm that, for the smallest $M=N$ tries to find a factorization up to $\varepsilon$ and, if failed, increments $M$ by $1$ and continues.
\begin{remark} In our computational experiments, we never had to go beyond $M = N$.
\end{remark}

Let $\mathbb{R}$ and $\mathbb{C}$ denote the fields of real and complex numbers, respectively. For a field $F$, such as $\mathbb{R}$ or $\mathbb{C}$, let $\GL(N,F)$ denote the group of all invertible $N\times N$ matrices with entries in $F$. 

\begin{lemma}
Let \[T_N := \{e^{i\diag(\phi_1,\ldots,\phi_N)}\mid \phi_1,\ldots,\phi_N \in \mathbb{R}\}\]
and, for a matrix $H$, 
\[G_H := \overline{\left\{e^{i\ell H}\mid \ell \in \mathbb{R}\right\}},\]
the closure as a real Lie group. Then, there exists a non-negative integer $M$ such that
\begin{equation}\label{eq:G}
G:=T_N\cdot G_H\cdot T_N\cdot G_H\cdot\ldots\ \text{($M$ times)},
\end{equation} is a real Lie subgroup of $U(N)$. Moreover, if the ratios of $H$'s non-zero eigenvalues are rational numbers, then
\[G_H = \left\{e^{i\ell H}\mid \ell \in \mathbb{R}\right\}.\]
\end{lemma}
\begin{proof}
By \cite[page~8, Problem~14]{VO1990}, the ratios of the non-zero elements from $a_1,\ldots,a_N$ are rational numbers if and only if \[D := \left\{e^{i\ell\diag(a_1,\ldots,a_N)}\mid \ell \in \mathbb{R}\right\}\] is a real Lie subgroup of $U(N)$. Therefore, the ratios of the non-zero eigenvalues of the matrix $H$ are rational if and only if $\left\{e^{i\ell H}\mid \ell \in \mathbb{R}\right\}$ is a real Lie subgroup of $U(N)$ and so $\left\{e^{i\ell H}\mid \ell \in \mathbb{R}\right\} = \overline{\left\{e^{i\ell H}\mid \ell \in \mathbb{R}\right\}} = G_H$. The result in~\eqref{eq:G} now follows from \cite[Section~7.5]{H1975}.
\end{proof}

Let $\mathcal{J}$ be the set of matrices $J$ such that the Lie algebra generated by $J$ and the Lie algebra of $T_N$ contains a matrix $\mathcal{M}$ that is conjugate to a single Jordan block of size $N$.

\begin{example}Consider a matrix $H$ from~\eqref{eq:Hmatrix} such that $\kappa_i\ne 0$ for all $i\in\{1,\ldots,N-1\}$. A direct computation shows that the vector space spanned over $\mathbb{C}$ by
\begin{itemize}
\item the Lie brackets of $H$ with the standard basis of the Lie algebra of $T_N$ and 
\item the Lie brackets of the above again with the standard basis of the Lie algebra of $T_N$
\end{itemize}
contains a matrix $\mathcal{M}$ that is conjugate to a single Jordan block of size $N$.
Therefore, $H \in \mathcal{J}$.
\end{example}

\begin{lemma} For all $H \in \mathcal{J}$,  the real Lie group $G$ defined by~\eqref{eq:G} is equal to  $U(N)$.
\end{lemma}
\begin{proof}
Since $G$ is a compact Lie group, its complexification $G_\mathbb{C}$ is a  reductive group \cite[page~246, Problem~22]{VO1990}. Moreover, $G_\mathbb{C}$ also contains a maximal torus $T_\mathbb{C}$ (the complexification of $T_N$) of $\GL(N,\mathbb{C})$ and so $G_\mathbb{C}$ is of maximal rank.
By \cite[Section~26.2, Corollary~A(b)]{H1975}, the center $Z(G_\mathbb{C})$ of $G_\mathbb{C}$ is contained in $T_\mathbb{C}$. 
By~\cite[Th\'eor\`eme~5]{BDS1949}, $G_\mathbb{C}$ is the connected component of the normalizer of and therefore is the centralizer of $Z(G_\mathbb{C})\subset T_\mathbb{C}$ in $\GL(N,\mathbb{C})$. Thus, $G_\mathbb{C}$ is conjugate to a direct product $\GL(k_1,\mathbb{C})\times\ldots\times \GL(k_s,\mathbb{C})$ for some positive integers $s$, $k_1,\ldots,k_s$ (see, e.g., \cite[Lemma~14.0.6]{G2011} or proof of \cite[Section 9.1, Proposition]{H1975}). Since the Lie algebra of $G_\mathbb{C}$ contains $\mathcal{M}$ that is conjugate to a single Jordan block of size $N$, we have: $s=1$ and $k_1\equiv k = N$. Hence, $G_\mathbb{C} = \GL(N,\mathbb{C})$ and so $G = U(N)$.
\end{proof}

\begin{corollary} Consider a matrix $H$ from~\eqref{eq:Hmatrix} such that $\kappa_i\ne 0$ for all $i\in\{1,\ldots,N-1\}$. Then, for every positive integer $N$, matrix $\mathcal{U} \in U(N)$, and $\varepsilon >0$, there exist a positive integer $M$ and sets of real numbers $\{\ell_p\}_{p=1}^{M-1}$ and $\{\phi_{n}^{(p)}\}_{p=1,n=1}^{N,M}$ forming the tuple $\boldsymbol{x}$ such that
\[\| \mathcal{U}-  \mathcal{U}(\boldsymbol{x})\| \leqslant\varepsilon\]  in the standard matrix norm (see notation~\eqref{eq_U}), where $\Vert A \Vert:=\sqrt{tr(AA^{\dagger})}$ is the Frobenius norm.
Moreover, if the ratios of  $H$'s non-zero eigenvalues are rational numbers, then $\boldsymbol{x}$ can be chosen so that $\mathcal{U}(\boldsymbol{x}) = \mathcal{U}$.
\end{corollary}

\section{Photonic Design and Performance}
\label{sec:phot}
\subsection{Practical realizability}
The proposed factorization is particularly suitable for optical implementations due to the nature of diagonal phase matrices, which can be represented in photonic circuits through layers of phase shifters. Recent progress on phase elements allows the tuning of phases of about $\pi$ radians in structures as long as 20 $\mu m$ using phase-change materials~\cite{Rios2022ultra,youngblood2023integrated}. In turn, the interlacing matrices $F_{\ell}=e^{i\ell H}$ match the wave evolution of electromagnetic waves through coupled waveguide arrays. 

This relation arises from coupled-mode theory~\cite{Huang94}, where the wave evolution is ruled by the discretized Schr\"odinger-like equation $-id\vec{E}/dz=H\vec{E}$, where the propagation coordinate $z$ plays the role of the time parameter in the Schr\"odinger equation. Here, $H$ is an $N\times N$ three-diagonal Hamiltonian characterizing the coupling between a given waveguide and its two nearest neighbours, related to the overlap of the evanescent waves outside the waveguide cores. 

Such an overlap is isotropic from one waveguide to its neighbour, leading to coupling parameters that render a Hermitian matrix Hamiltonian $H$. In this form, the tridiagonal matrix form used in our proof is compatible with the matrix representation of coupled waveguide arrays. Two noteworthy cases for $H$ are the $J_x$ lattice~\cite{Ata97,weimann2016implementation,honari_2020} and the homogeneous lattice~\cite{christodoulides2003discretizing,guerrero2021coherent}, both with on-site terms $\nu_{n}=0$ and coupling parameters \[\kappa^{(Jx)}_{n}=\sqrt{(N-n)n}/2\quad \text{and}\quad \kappa^{(h)}_{n}=1 ,\] respectively, for $n\in\{1,\ldots,N-1\}$. The former has integer eigenvalues \[\lambda^{(Jx)}_{m}\in\{n\}_{n=-j}^{j},\quad  N=2j+1,\] whereas the second has \[\lambda^{(h)}_{m}=2\cos(\pi m/(N+1)), \quad m\in\{1,\ldots,N\}.\] These are examples of Hamiltonians with either rational and irrational eigenvalue ratios so that an arbitrary $\varepsilon$ precision can be achieved (see Corollary II.1).

\begin{figure}
\flushleft
\includegraphics[width=0.45\textwidth]{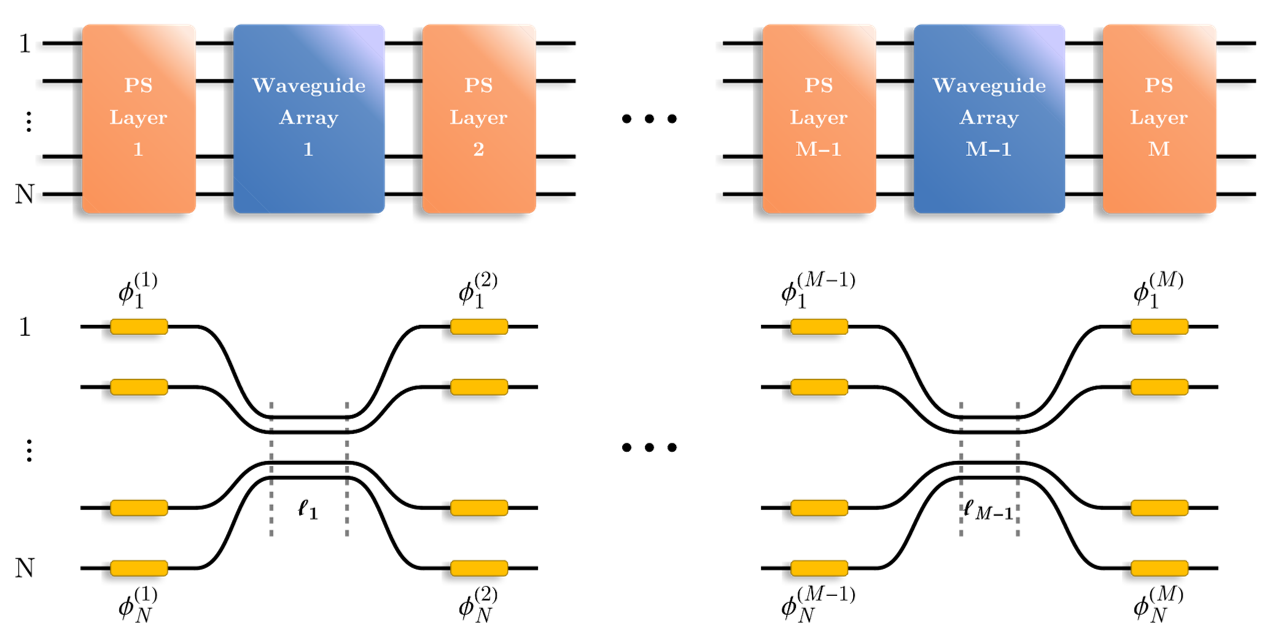}
\caption{Sketch of the proposed factorization and its physical realization for a $N$-port device through waveguide arrays with lengths $\ell_{j}$ and phase elements $\phi_{n}^{(m)}$ for a finite number of layers $M$.}
\label{fig1}
\end{figure}

\begin{remark}
Although our proof has established the existence of a finite number of layers $M$ to render the universality of the proposed factorization, the exact value of $M$ is yet to be found. Furthermore, for an arbitrary $N\times N$ unitary matrix $\mathcal{U}$, our factorization yields $M(N+1)-1$ free parameters for a $M$-layer architecture, whereas any matrix $\mathcal{U}\in U(N)$ is defined, in general, by $N^2$ parameters. This establishes the lower bound for the number of layers as $M\geq N$. 
\end{remark}

\subsection{Parameter optimization}
Let $\mathcal{U}_{t}$ be an arbitrary unitary target matrix and \[\boldsymbol{x}=\{\phi_{p}^{m}\}_{p=1,m=1}^{N,M}\cup\{\ell_{m}\}_{m=1}^{M-1}\] the tuple of parameters used in the factorization~\eqref{eq_U} to reconstruct the target matrix. To determine the number of layers $M$ and the parameter set $\boldsymbol{x}$, we define the error norm as the positive semi-definite function
\begin{equation}
L(\boldsymbol{x})=\frac{\Vert \mathcal{U}(\boldsymbol{x})-\mathcal{U}_t \Vert^{2}}{N^{2}}.
\label{eq_L}
\end{equation}
If, for some $\boldsymbol{x}$, the error norm minimum is zero, $L(\boldsymbol{x})=0$, the factorization~\eqref{eq_U} exactly reconstructs the target matrix $\mathcal{U}_{t}$. Since $M=N$ is the minimum number of layers (equivalently $N^{2}+N-1$ parameters), the factorization $\mathcal{U}(\boldsymbol{x})$ poses an overparamaterized problem, the solution of which ($\boldsymbol{x}$) may not be unique. Thus, any set $\boldsymbol{x}$ that produces an error norm below a prescribed tolerance is accepted as a solution. For this task, we focus on the Levenberg-Marquardt algorithm, suitable for least-square problems such as the one defined by the error norm~\eqref{eq_L}, proved useful in other similar optimization tasks~\cite{Markowitz23,Markowitz2023auto} 


\begin{remark} Because of the numerical nature of the optimization routine, no set of parameters renders $L$ exactly zero. Thus, one has to impose a performance tolerance $\delta>0$ so that the parameter set $\boldsymbol{x}$ is said to reconstruct $\mathcal{U}_t$ whenever $L\leq \delta$.
\end{remark}

\begin{figure}
\centering
\includegraphics[width=0.45\textwidth]{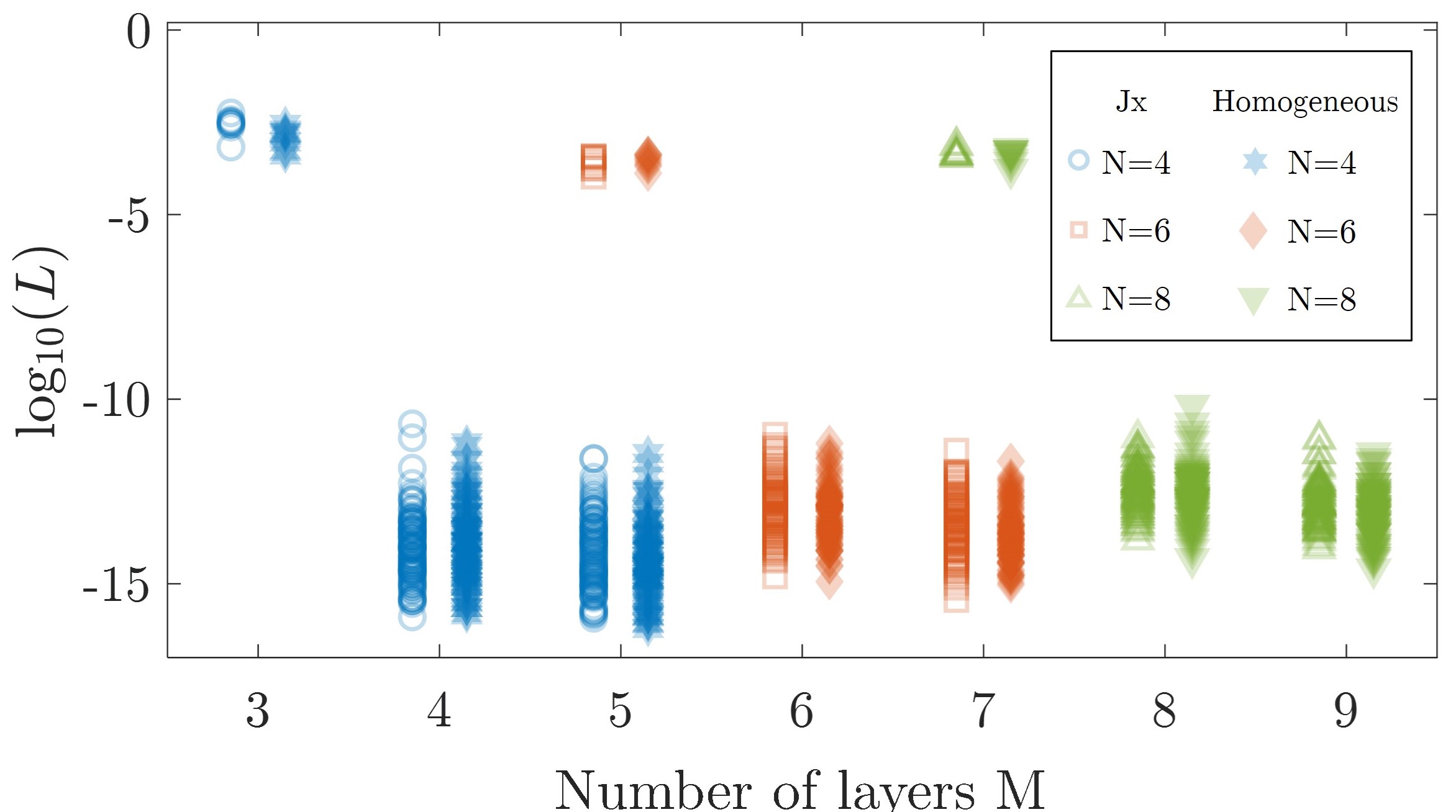}
\caption{Error norm log${}_{10}(L)$ obtained from the optimization routine for the $J_x$ (open markers) and homogeneous (filled markers) lattice. For both cases, the matrix dimension has been fixed to $N=4,6,8$, whereas the number of layers for each case is $M=N-1$, $M=N$, $M=N+1$.}
\label{Fig2}
\end{figure}

To avoid any bias on the choice of target matrices, we randomly construct $N_T$ target matrices generated through the Haar measure~\cite{mezzadri_how_2007}. The error norm for each target matrix is optimized by fixing $M$ and running the optimization routine until either $L\leq \delta$ or $N_{it}$ iterations have been performed. The Hamiltonians $H$ used for testing purposes correspond to the $J_x$ and the homogeneous lattices. Results of the optimization routine are presented in Figure~\ref{Fig2}, with tests run for $N=4,6,8$ and $M=N-1,N,N+1$. Although it was previously established that $M\geq N$, we perform the optimization for $M=N-1$ to identify any transition in the behavior of $L$. $N_T=100$ target matrices were tested for each configuration. In all testing cases, the error norm drastically drops at $M=N$, and no further improvement is found for $M=N+1$. This reveals that such transition indeed occurs for $M=N$, which was set as the required minimum number of layers to achieve universality for~\eqref{eq_U}. 

\subsection{All-optical logic gate circuit and simulation}
\begin{figure*}
\centering
\includegraphics[width=0.95\textwidth]{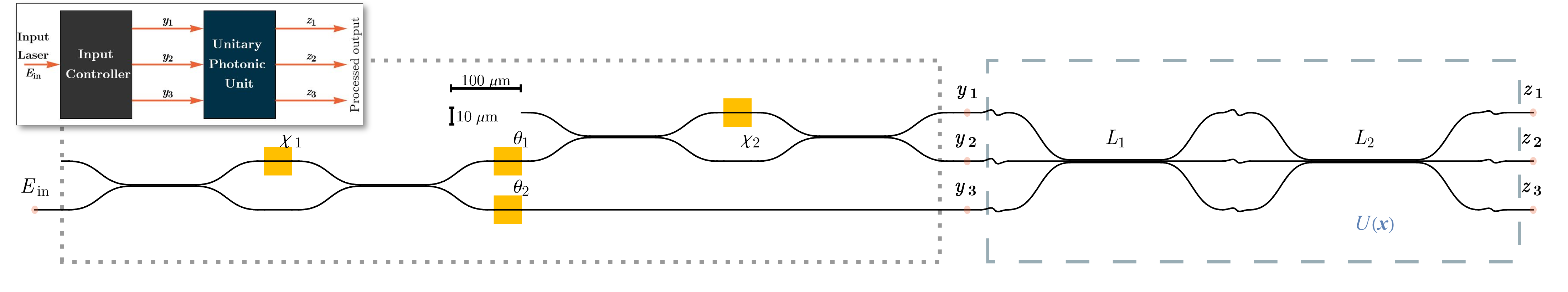}
\caption{Block diagram and photonic circuit sketch for the proposed logic gate operation. A single input field $E_{in}$ excites the architecture, producing the logic gate input fields $y_{1,2,3}$ during the input controller stage (dotted rectangle) by tuning the active phase shifters $\chi_{1}$, $\chi_{2}$, $\theta_{1}$, and $\theta_{2}$ (yellow boxes). The latter inputs are then processed in the photonic unitary unit constructed through Eq.~\eqref{eq_U} (dashed rectangle). Here, the physical waveguide coupling lengths are $L_{j}=\widetilde{\kappa}\ell_{j}$, with $\widetilde{\kappa}$ a design scaling factor and $\ell_{j}$ the optimized length parameters.}
\label{Fig3}
\end{figure*}
The proposed architecture is examined by considering $N=3$ ports so that eleven parameters shall be optimized; that is, nine phase shifters and two waveguide lengths. Furthermore, the $J_x$ lattice Hamiltonian is used as the exponential Lie algebra generator $H$. This particular choice is handy since its eigenvalues $\lambda_{n}^{(J_x)}=n$, rendering a unitary evolution that is periodic $e^{i(\ell+2\pi)H}\equiv e^{i\ell H}$. That is, the lattice length becomes $ \ell \mod 2\pi$. Thus, even if the optimization routine produces a negative or a large value of $\ell$, we can always find the equivalent one $\ell\in(0,2\pi]$. In this form, we do not need to constrain the optimization routine, which speeds up the optimization process.

As a particular example, we consider constructing an all-logical logic gate device. Indeed, such a task has been achieved on combined gain and loss in an optical fiber~\cite{williams2013all} and stub optical lattices showcasing a flat band in the dispersion relations~\cite{real2017flat}. In the former, a non-unitary process is required to account for a lossy system, whereas the second approach involves a multilayer design so that the decorated waveguides can be incorporated into the photonic lattice. 

The proposed approach achieves the logic gate operation using a lossless optical circuit design. To this end, we consider the non-canonical orthonormal basis
\[ \vec{v}_{1}=\frac{(1,-1,0)}{\sqrt{2}}, \quad \vec{v}_{2}=\frac{(1,1,-\sqrt{2})}{2}, \quad \vec{v}_{3}=\frac{(1,1,\sqrt{2})}{2}\] 
so that the target unitary becomes $\mathcal{U}_{t}=(\vec{v}_{1}^{T},\vec{v}_{2}^{T},\vec{v}_{3}^{T})$. The corresponding reconstructed matrix $\mathcal{U}(\boldsymbol{x})$ is represented through~\eqref{eq_U} after proper optimization of the involved parameter set $\boldsymbol{x}$.

The photonic circuit performing the logic gate operation is divided into two parts. In the first stage, a series of programmable Mach-Zehnder interferometers (MZI) controlled by the phase shifters $\chi_{1,2}$ and $\theta_{1,2}$ is introduced so that the input laser $E_{in}$ splits into the inputs $y_{1,2,3}$ used for the logic gate operation, where $y_i \in\mathbb{C}$, $1\leqslant i\leqslant 3$. See the dotted rectangle in Figure~\ref{Fig3}. Here, $y_{1,2}$ are the logic ports to be processed, whereas $y_{3}$ is used as a control port to control the logic gate operation. In the second stage, the photonic unit comprises the interlaced structure~\eqref{eq_U}, implemented through waveguide arrays, and passive phase elements. See the dashed rectangle in Figure~\ref{Fig3}. This unit processes the inputs $y_{1,2,3}$ into the required logic operation at the output ports $z_{1,2,3}$.

\begin{table*}[]
    \centering
    \begin{tabular}{c|ccc|ccc}
     $(\chi_1,\chi_2)$ & $\vert y_1\vert^2$ & $\vert y_2\vert^2$ & $\vert y_3\vert^2$ & $\vert z_1\vert^2$ & $\vert z_2\vert^2$ & $\vert z_3\vert^2$  \\ 
     \hline
     No $E_{in}$ & 0  & 0  &  0 &  0 & 0  & 0 \\
     $(0,0)$ & 1  & 0 &  0 & $\frac{1}{2}$  & $\frac{1}{4}$  & $\frac{1}{4}$ \\
     $(0,\pi)$ & 0 & 1 & 0 & $\frac{1}{2}$ & $\frac{1}{4}$ & $\frac{1}{4}$ \\
     $\left(0,\frac{\pi}{2}\right)$ & $\frac{1}{2}$ & $\frac{1}{2}$ & 0 & 0 & $\frac{1}{2}$ & $\frac{1}{2}$ \\
     \hline
     & & & & \texttt{XOR} & \texttt{OR} & \texttt{OR} \\
     & & & & \footnotesize{$Th_{1}<\frac{1}{2}$} & \footnotesize{$Th_{2}<\frac{1}{4}$} & \footnotesize{$Th_{3}<\frac{1}{4}$} 
    \end{tabular}
    \quad
    \begin{tabular}{c|ccc|ccc}
     $(\chi_1,\chi_2)$ & $\vert y_1\vert^2$ & $\vert y_2\vert^2$ & $\vert y_3\vert^2$ & $\vert z_1\vert^2$ & $\vert z_2\vert^2$ & $\vert z_3\vert^2$  \\ 
     \hline
     $(\pi,\textnormal{any})$ & 0  & 0  &  1 &  0 & $\frac{1}{2}$  & $\frac{1}{2}$ \\
     $(\tilde{\chi}_{0},0)$ & $\frac{1}{3}$  & 0 &  $\frac{2}{3}$ & $\frac{1}{6}$  & $\frac{1}{12}$  & $\frac{9}{12}$ \\
     $(\tilde{\chi}_{0},\pi)$ & 0 & $\frac{1}{3}$ & $\frac{2}{3}$ & $\frac{1}{6}$ & $\frac{1}{12}$ & $\frac{9}{12}$ \\
     $\left(\frac{\pi}{2},\frac{\pi}{2}\right)$ & $\frac{1}{4}$ & $\frac{1}{4}$ & $\frac{1}{2}$ & 0 & 0 & 1 \\
     \hline
     & & & & \texttt{XOR} & \texttt{NAND} & \texttt{AND} \\
     & & & & \footnotesize{$Th_{1}<\frac{1}{6}$} & \footnotesize{$Th_{2}<\frac{1}{12}$} & \footnotesize{$\frac{9}{12}<Th_{3}<1$} 
    \end{tabular}
    \caption{Photonic unit normalized power inputs ($\vert y_{j}\vert^{2}$) and power outputs ($\vert z_{j}\vert^{2}$), for $j\in\{1,2,3\}$, used to perform the logic gate operation. The operation settings are shown for different configurations of the tunable phase duple $(\chi_1,\chi_2)$, where $\widetilde{\chi}_{0}=2\textnormal{arctan}(\sqrt{2})$.}
    \label{tab:Table1}
\end{table*}

Each interlacing unitary matrix is implemented through coupled waveguide arrays of different lengths. The waveguides are built up on a silicon-on-silica structure, where the core (Si) and the surrounding silica cladding (SiO$_2$) refractive indexes are $n_{co}=3.41$ and $n_{cla}=1.4711$, respectively. More specifically, the waveguide cores have a 500 nm width and 200 nm height, surrounded by a silica substrate. This enables the generation of only the fundamental TE mode. Full-wave simulations are computationally expensive to implement; we thus approximate the 3D waveguide system to an equivalent 2D model that carries the same information as the 3D one. This is done using \textit{effective index theory}, from which one obtains the parameters $n_{c}^{(\textnormal{eff})}=2.71$ for a monochromatic input source at 1550 nm. An array of Mach-Zehnder interferometers is coupled to the architecture input to individually manipulate the input ports $y_{1,2,3}$ from a single source. See Appendix~\ref{appendix} for details.

The complete integrated photonic circuit used in the simulations is sketched in Figure~\ref{Fig3}. Here $L_{1,2}=\widetilde{\kappa}\ell_{1,2}$ is the physical lattice coupling length, with $\widetilde{\kappa}\approx (280/\pi)$ $\mu m/ rad$ and $\ell_{1,2}$ the lengths obtained from the optimization routine. Note that the first stage of the architecture has a series of MZIs that split the input field $E_{in}$ in the required input ports $y_{1,2,3}$. Each MZI comprises two 50/50 directional couplers with coupling lengths of 96 $\mu m$ and separated by 500 nm (edge-to-edge distance). Likewise, the waveguide array separation (edge-to-edge) is set to 500 nm, whereas the coupling lengths $L_{j}$ are determined once the optimization is performed.

The phase shifters $\chi_{1,2}$ allow for tuning the desired input in $\mathcal{U}_{t}$, whereas the phase shifters $\theta_{1,2}$ are introduced to compensate for any phase change produced during the amplitude modulation. Tunable phase elements can be introduced using phase-change materials~\cite{Rios2022ultra}, where a phase modulation as compact as $11\mu m$ per $\pi$ radians is feasible. In turn, waveguide rerouting is used to introduce the passive phase delays $\phi_{j}^{(n)}$ extracted from the optimization. 

\begin{figure*}
\centering
\includegraphics[width=0.92\textwidth]{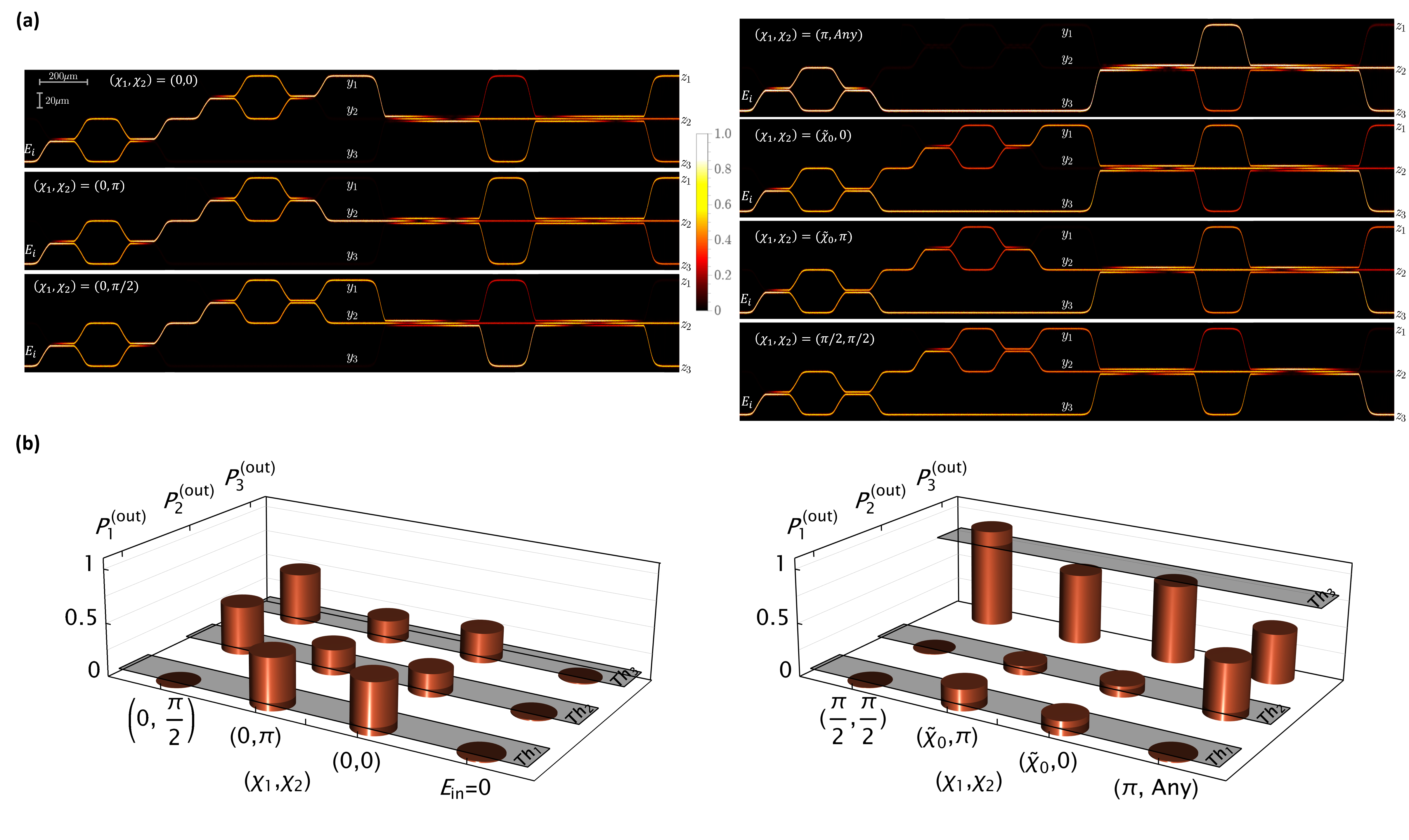}
\caption{(a) Density plot of $\vert \vec{E}(x,y)\vert$ computed from full-wave simulations for different operation settings of the duple $(\chi_1,\chi_2)$. (b) Normalized output power $P^{(out)}_{j}\propto \vert z_{j} \vert^{2}$ for the configurations in Table.~\ref{tab:Table1} and the corresponding threshold values $Th_j$, for $j\in\{1,2,3\}$.}
\label{Fig4}
\end{figure*}

The phase shifters are operated according to the rules shown in Table~\ref{tab:Table1} to produce the required input and output operation. The control port $y_3$ is accordingly manipulated to switch the logic gate operation. For $y_3=0$ (Table~\ref{tab:Table1}(a)), the \texttt{XOR} and \texttt{OR} operations are reproduced at the ports $z_{1}$ and $z_{2,3}$, respectively. A second operation is established for $y_{3}\neq 0$ (Table~\ref{tab:Table1}(b)), where the logic gates \texttt{NAND} and \texttt{AND} are now performed in the ports $z_2$ and $z_3$, respectively, as long as the threshold values are correctly fixed. We particularly focus on the case $y_3=\sqrt{2}y_2$, with $y_2\neq 0$, so that $z_3=0$ for $y_1=y_2$, which is required to perform the \texttt{XOR} and \texttt{NAND} gates. This particular case is possible by tuning $(\chi_{1},\chi_{2})=(\widetilde{\chi}_{0},\pi)$ and $(\chi_{1},\chi_{2})=(\pi/2,\pi/2)$, where $\widetilde{\chi}_{0}=\textnormal{arctan}(\sqrt{2})$. 

To illustrate the functionality of the proposed architecture, the corresponding full-wave simulations are shown in Figure~\ref{Fig4}(a) for each of the settings presented in Table~\ref{tab:Table1}. Here, the simulated power distribution throughout the device is portrayed for all the settings, where the unitary unit input and output ports $y_{j}$ and $z_{j}$, respectively, so that the proper operation can be tracked. This shows the light routing for all the phase shifter configurations. For this particular device, the phase information is irrelevant. Thus, no further measurements other than the output power $P^{(out)}_{j}\propto \vert z_{j}\vert^2$ are required. 

In contradistinction to binary operations, the measured optical power $P^{(out)}_{j}\propto \vert z_{j}\vert^{2}$ are positive semidefinite quantities. Thus, to establish a binary-like operation, we define the threshold values $Th_j$ for each output port $z_j$ so that $P^{(out)}_{j}<Th_j$ and $P^{(out)}_{j}\geq Th_j$ can be interpreted as logic 0 and 1 readouts, respectively, for $j\in\{1,2,3\}$. Furthermore, the threshold values must be fixed to account for deviations due to mild manufacturing imperfections and design flaws. The ideal thresholds shown in Table~\ref{tab:Table1} provide us with the allowed threshold intervals. Indeed, Figure~\ref{Fig4}(b) shows the simulated measurement for all the operation settings. The horizontal strips denote the pre-fixed normalized threshold values $Th_{j}$, fixed to recover the desired logic gate functionality and to account for deviations from the ideal case due to channel cross-talk, mild defects, and losses around the bending arms. For instance, in the case $(\pi/2,\pi/2)$, no power is expected at the ports $z_{1,2}$, yet a small amount is observed in the simulations. The proper readouts are gathered by fixing the thresholds $Th_{1}=Th_{2}=0.075$, whereas $Th_{3}=0.8$ allows recovering the \texttt{AND} operation at the port $z_{j}$ for $y_3\neq 0$.

\section{Conclusions}
The present work provides a formal proof for the universality of photonic interlacing architectures in representing the discrete linear unitary group $U(N)$. It demonstrates that elements of $U(N)$ can be decomposed into a sequence of $N$-parameter phases alternating with $1$-parameter propagators of a lattice Hamiltonian, with a finite number of layers. Numerical tests using Haar random unitary matrices as targets reveal that specifically $N$ layers are required to reconstruct the targets with error norms in the numerical error regime. 

The proposed architecture can be implemented using basic photonic circuit components and is capable of reconstructing arbitrary unitary matrices. Unlike other similar structures, this solution requires one less layer, resulting in an $M=N$ layered design instead of $M=N + 1$. This reduction is feasible because we achieve the minimum number of parameters needed to cover the \( U(N) \) group, which is $N^2-1$ if we account for one global phase. In our design, we can factor out a global phase for each phase layer, allowing for $N-1$ parameters per phase layer. Additionally, the length parameter $\ell_{j}$ is also trainable, yielding a total of $N$ parameters per layer combination $e^{i\ell_{p}}F$, with $p\in\{1,\ldots,N\}$.

Numerical tests and full-wave simulations corroborate the theoretical findings, confirming the potential of this approach for developing energy-efficient and compact on-chip photonic devices for various applications. Among the potential applications of the present architecture, we demonstrate the feasibility of designing an all-logical logic gate device using a lossless and passive optical circuit compatible with open-access silicon foundries. Nevertheless, the proposed architecture is not limited to logic gate operations and can be deployed for other optical signal processing tasks that require unitary operations.

\begin{acknowledgments}
We are grateful to Andrei Minchenko for his useful suggestions. This project is supported by the U.S. Air Force Office of Scientific Research (AFOSR) Young Investigator Program (YIP) Award\# FA9550-22-1-0189, the City University of New York (CUNY) Junior Faculty Research Award in Science and Engineering (JFRASE) funded by the Alfred P. Sloan Foundation, and the National Science Foundation Award\# CNS-2329021.
\end{acknowledgments}
\noindent
\noindent

\appendix
\section{}
\label{appendix}
The circuit design presented in the main manuscript is divided into two parts: the programmable MZI array that produces the desired input signal and the passive photonic unit that processes the input and produces the logic gate operation at the output. In the current design, each MZI is composed of two passive directional couplers and two active phase shifters (yellow boxes in Figure~\ref{Fig3}). Each of the latter components is described through a unitary matrix. Particularly, the wave evolution in each directional coupler is determined by the coupled-mode theory, leading to \[U_{\texttt{DC}}=e^{-i \ell \kappa \sigma_{1}} \equiv \cos(\kappa \ell) \sigma_{o}-i\sin(\kappa \ell)\sigma_{1},\] with $\sigma_{0}$ the $2\times 2$ identity matrix and $\sigma_{1}\equiv\sigma_{x}$ the corresponding x-projection of the Pauli matrix. In the latter, it is assumed that both waveguides have the same dimensions so that any additional phase mismatch can be disregarded, and $\kappa$ is proportional to the overlap integral of the guided modes in each waveguide~\cite{yariv2007photonics}. Here, the waveguide width and thickness are 500 nm and 220 nm, respectively. The edge-to-edge distance between waveguides has been fixed to 500 nm, which, combined with a coupling length of 96 $\mu m$, renders a 50/50 or power divider operation ($\kappa \ell=\pi/4$). Additionally, the phase shifters layers are represented by diagonal matrices, and the input vector is fixed as $(0,0,E_{\texttt{in}})^{T}$, where the input field enters the lower-most channel (third input). From the previous considerations and following the circuit shown in Figure~\ref{Fig3}, it is straightforward to show that the photonic unit inputs take the form 
\begin{equation*}
\begin{aligned}
& y_{1}=e^{i\left( \theta_{1}+\frac{\chi_1+\chi_2}{2}-\pi\right)}\cos\left(\frac{\chi_{1}}{2}\right)\sin\left(\frac{\chi_2}{2}\right), \\
& y_{2}=e^{i\left( \theta_{1}+\frac{\chi_1+\chi_2}{2}-\pi\right)}\cos\left(\frac{\chi_{1}}{2}\right)\cos\left(\frac{\chi_2}{2}\right), \\
& y_{3}=e^{i\left( \theta_{2}+\frac{\chi_1}{2}-\frac{\pi}{2}\right)}\sin\left(\frac{\chi_{1}}{2}\right). \\
\end{aligned}
\end{equation*}
It is worth noting that the phase shifters $\chi_{1,2}$ produce amplitude modulation at $y_{1,2,3}$ while introducing additional phases. To address this, the extra phase shifters $\theta_{1,2}$ were included to compensate for any other phase change produced during the modulation or waveguide routing. Thus, the elements $\theta_{1,2}$ are a function of $\chi_{1,2}$, which we fix as $\theta_{1}=(\pi-(\chi_1+\chi_2)/2)$ and $\theta_{2}=(\pi-\chi_1)/2$ to remove any extra phase. From this, the results shown in Table~\ref{tab:Table1} follow straightforwardly.

\bibliography{biblio} 

\end{document}